\documentclass{article}\usepackage[]{graphicx}\usepackage[]{color}
\makeatletter
\def\maxwidth{ %
  \ifdim\Gin@nat@width>\linewidth
    \linewidth
  \else
    \Gin@nat@width
  \fi
}
\makeatother

\definecolor{fgcolor}{rgb}{0.196, 0.196, 0.196}

\usepackage{framed}
\makeatletter
\newenvironment{kframe}{%
 \def\at@end@of@kframe{}%
 \ifinner\ifhmode%
  \def\at@end@of@kframe{\end{minipage}}%
  \begin{minipage}{\columnwidth}%
 \fi\fi%
 \def\FrameCommand##1{\hskip\@totalleftmargin \hskip-\fboxsep
 \colorbox{shadecolor}{##1}\hskip-\fboxsep
     \hskip-\linewidth \hskip-\@totalleftmargin \hskip\columnwidth}%
 \MakeFramed {\advance\hsize-\width
   \@totalleftmargin\z@ \linewidth\hsize
   \@setminipage}}%
 {\par\unskip\endMakeFramed%
 \at@end@of@kframe}
\makeatother

\definecolor{shadecolor}{rgb}{.97, .97, .97}
\definecolor{messagecolor}{rgb}{0, 0, 0}
\definecolor{warningcolor}{rgb}{1, 0, 1}
\definecolor{errorcolor}{rgb}{1, 0, 0}
\newenvironment{knitrout}{}{} 

\usepackage{alltt}
\usepackage[italian,english]{babel}
\usepackage[margin=3cm]{geometry}
\usepackage{natbib}
\usepackage{float}
\usepackage{multirow} 
\usepackage{url} 
\usepackage{bm} 
\usepackage{amsmath}
\usepackage[font=small, labelfont=bf]{caption}
\usepackage[doublespacing]{setspace}
\usepackage{lineno}

\providecommand{\keywords}[1]
{
  \small	
  \textbf{\textit{Keywords---}} #1
}

\usepackage{hyperref}

\title{\huge Enhancing statistical inference\\ in psychological research via prospective\\ and retrospective design analysis}

\author{Gianmarco Alto\`e\,$^{1}$\thanks{Correspondence:\newline
Gianmarco Alto\`e, Department of Developmental Psychology and Socialization,\newline
University of Padova, Via Venezia 8, 35131 Padova, Italy\newline
\href{mailto:gianmarco.altoe@unipd.it}{gianmarco.altoe@unipd.it}}, Giulia Bertoldo\,$^{1}$, Claudio Zandonella Callegher\,$^{1}$,\\ Enrico Toffalini\,$^{2}$, Antonio Calcagn\`i\,$^{1}$, Livio Finos\,$^{1}$ and Massimiliano Pastore\,$^{1}$ }
\date{\small  $^1$ Department of Developmental Psychology and Socialisation, University of Padova, Padova, Italy\\
\small $^2$ Department of General Psychology, University of Padova, Padova, Italy}

\IfFileExists{upquote.sty}{\usepackage{upquote}}{}
\begin{document}

\maketitle

\noindent\rule{\linewidth}{0.6pt}

\begin{abstract}
In the past two decades, psychological science has experienced an unprecedented replicability crisis
which uncovered several problematic issues. Among others, the use and misuse of statistical inference plays a key role in this crisis. Indeed, statistical inference is too often viewed as an isolated procedure limited to the analysis of data that have already been collected. Instead, statistical reasoning is necessary both at the planning stage and when interpreting the results of a research project.  Based on these considerations, we build on and further develop an idea proposed by Gelman and Carlin (2014) termed ``prospective and retrospective design analysis''. Rather than focusing only on the statistical significance of a result and on the classical control of type I and type II errors, a comprehensive design analysis involves reasoning about what can be considered a plausible effect size. Furthermore, it introduces two relevant inferential risks: the exaggeration ratio or Type $M$ error (i.e., the predictable average overestimation of an effect that emerges as statistically significant), and the sign error or Type $S$ error (i.e., the risk that a statistically significant effect is estimated in the wrong direction). Another important aspect of design analysis is that it can be usefully carried out both in the planning phase of a study and for the evaluation of studies that have already been conducted, thus increasing researchers’ awareness during all phases of a research project. To illustrate the benefits of design analysis to the widest possible audience, we use a familiar example in psychology where the researcher is interested in analyzing the differences between two independent groups considering Cohen’s d as an effect size measure. We examine the case in which the plausible effect size is formalized as a single value, and propose a method in which uncertainty concerning the magnitude of the effect is formalized via probability distributions. Through several examples and an application to a real case study, we show that even though a design analysis requires big effort, it has the potential to contribute to planning more robust and replicable studies. Finally, future developments in the Bayesian framework are discussed.
\end{abstract} \hspace{10pt}


\noindent\rule{\linewidth}{0.6pt}



  \begin{quotation}
  \begin{flushright}
    \textit{``If statisticians agree on one thing, it is that\\ scientific inference should not be made mechanically.''\\
    \citet[p. 422]{gigerenzer2015surrogate} }
    \end{flushright}
  \end{quotation}
  \begin{quotation}
  \begin{flushright}
      \textit{``\textbf{A}ccept uncertainty. Be \textbf{t}houghtful, \textbf{o}pen, and \textbf{m}odest.\\ Remember \lq ATOM\rq.'' \\ \citet[p. 2]{wasserstein2019moving}}  
   \end{flushright}
   \end{quotation}

\section{Introduction}
In the past two decades, psychological science has experienced an unprecedented replicability
crisis \citep{ioannidis2005most, pashler2012editors, opensciencecollaborationEstimatingReproducibilityPsychological2015a} which uncovered a number of problematic issues, including the adoption of Questionable Research Practices \citep{john2012measuring} and Questionable Measurement Practices \citep{flake2019measurement}, the reliance on excessively small samples \citep{button2013a}, the misuse of statistical techniques \citep{pastore2019potenza}, and the consequent misleading interpretation and communication of research findings \citep{wasserstein2019moving}.

Whereas some important reasons for the crisis are intrinsically related to psychology as a science \citep{chambers2019seven}, which lead to a renewed recommendation to rely on strong and well-formalized theories when planning a study, the use of statistical inference undoubtedly plays a key role. Specifically, the inferential approach most widely used in psychological research, namely Null Hypothesis Significance Testing (NHST), has been strongly criticized \citep{gigerenzer2004null, gelman2018failure, mcshane2019abandon}. As a consequence, several alternative approaches have received increasing attention, such as the use of Bayes Factors for hypotesing testing, and the use of both Frequentist and Bayesian methods to estimate the magnitude of the effect of interest with uncertainty (see \citeauthor{kruschke2018bayesian}, \citeyear{kruschke2018bayesian}, for a comprehensive historical review).

In the current paper we focus on an upstream, but still neglected issue that is unrelated to the approach chosen by the researcher, namely the need for statistical
reasoning, i.e., ``to reason about data, variation and chance'' \citep[p. 1253]{moore1998statistics}, during all phases of an empirical study. Our work was inspired by the famous statistician Ronald Fisher (1890 -1962), who stated that ``To consult the statistician after an experiment is finished is often merely to ask him to conduct a post mortem examination. He can perhaps say what the experiment died of'' \citep[p. 17]{fish}. Indeed, we argue that too often statistical inference is seen as an isolated procedure, which is limited to the analysis of data that have already been collected. In particular, we emphasize the non-trivial importance of making statistical considerations at the onset of a research project. Furthermore, we stress that although Fisher has ironically defined them a “post mortem examination”, appropriate evaluations of published results can provide a relevant contribution to the progress of (psychological) science. The ultimate goal of this paper is to increase researchers’ awareness by promoting active engagement when designing their research.

To achieve this goal, we build on and further develop an idea proposed by \citeauthor{gelmanPowerCalculationsAssessing2014a} (\citeyear{gelmanPowerCalculationsAssessing2014a}) called ``prospective and retrospective design analysis'', which is virtually absent in current
research practice. Specifically, to illustrate the benefits of design analysis to the widest possible audience, we use a familiar example in psychology  where the researcher is interested in analyzing the differences between two independent groups considering Cohen’s $d$ \citep{cohenStatisticalPowerAnalysis1988} as an effect size measure.

In brief, the term \textit{design analysis} has been proposed by \citeauthor{gelmanPowerCalculationsAssessing2014a} (\citeyear{gelmanPowerCalculationsAssessing2014a}) as a broader definition of power analysis,  a concept that in the statistical literature traditionally indicates the determination of an appropriate sample size given prespecified levels of Type I and Type II errors and a “plausible effects size” \citep{gigerenzer2004null}. Indeed, a comprehensive design analysis should  also explicitly consider other inferential risks, including the exaggeration ratio (Type $M$ error, i.e., the predictable average overestimation of an effect that emerges as statistically significant) and the sign error (Type $S$ error, i.e., the risk that a statistically significant effect is estimated in the wrong direction). Notably, the estimation of these errors will require an effort from psychologists to introduce their expert knowledge and hypothesize what could be considered a ``plausible effect size''. As we will see later, a key aspect of design analysis is that it can be usefully carried out both in the planning phase of a study (i.e., prospective design analysis) and for the evaluation of studies that have already been conducted (i.e., retrospective design analysis).

Although the idea of design analysis could be developed within different inferential statistical approaches (e.g., Frequentist and Bayesian), in this paper we will rely on the Neyman-Pearson (N-P) approach \citep{pearson1928use} as opposed to the widely used NHST. 
The rationale for this choice is that, in addition to other strengths, the N-P approach includes formalization of the \textit{Null Hypothesis} (i.e., the absence of an effect) like NHST, but also an explicit formalization of the \textit{Alternative Hypothesis} (i.e., the magnitude of the expected effect). For a more comprehensive description of the difference between N-P and NHST approaches, we refer the reader to Gigerenzer and colleagues (\citeyear{gigerenzer2004null}).

The remainder of this paper is structured as follows. In the next paragraphs, we will briefly review the main consequences of underpowered studies, discuss two relevant misconceptions concerning the interpretation of statistically significant results, and present a theoretical framework for design analysis including some clarifications regarding the concept of ``plausible effect size''. In Section \ref{pr_and}, through familiar examples within psychological research, the benefits of prospective and retrospective design analysis will be highlighted. In Section \ref{inc_unc}, we will propose a specific method that, by explicitly taking uncertainty issues into account, could further assist researchers in evaluating scientific findings. Subsequently, in Section \ref{an_ill}, a real case study will be presented and analyzed. Finally, in Section \ref{dis_and}, we will summarize potentials, further developments, and limitations of our proposal.

To increase readability and to ensure transparency of our work, we also include two appendices:
\begin{itemize}
\item \textbf{Appendix A}. A detailed description concerning the computation and the interpretation of Cohen's $d$.
\item \textbf{Appendix B}. A brief explanation of the ad-hoc R \citep{R2018} functions used in the paper. Details on how to reproduce the presented examples and on how to use our R functions for other purposes, are also provided. Furthermore, the source code of our functions, \texttt{functions\_PRDA.R},
is freely available at the Open Science Framework (OSF) at the link \url{https://osf.io/j8gsf/files/}.
\end{itemize}

\subsection{The consequences of underpowered studies in Psychology}

In 1962, Cohen raised attention towards a problem affecting psychological research that is still very much alive today \citep{cohen1962a}. Researchers seemed to ignore the statistical power of their studies - which is not considered in NHST  \citep{gigerenzer2004null} - with severe consequences for the robustness of their research findings.
In the N-P approach, the power of a statistical test is defined as the probability that the test has to reject the Null Hypothesis ($H_0$) when the Alternative Hypothesis ($H_1$) is true. One of the problems with underpowered studies is that the probability of finding an effect, if it actually exists, is low. More importantly, if a statistically significant result (i.e., ``in general'', when the observed $p-value$ is less than .05, and consequently $H_0$ is rejected; see, \citeauthor{wasserstein2019moving}, \citeyear{wasserstein2019moving}) is obtained in an underpowered study, the effect size associated with the observed $p-value$ might be ``too big to be true'' \citep{button2013a,gelmanPowerCalculationsAssessing2014a}.

This inflation of effect sizes can be seen when examining results of replication projects, which are usually planned to have higher power than the original studies. For example, the \citet[pp. 4-5]{opensciencecollaborationEstimatingReproducibilityPsychological2015a} reported that ``Overall, original study effect sizes ($M = 0.403$, $SD = 0.188$) were reliably larger than replication effect sizes ($M= 0.197$, $SD = 0.257$)'', and in the Social Science Replication Project \citep[p. 637]{camerer2018a}, ``the effect size of the replication was on average about 50\% of the original effect size''. These considerations contributed to the introduction in the literature of the term ``decline effect'', defined as ``the notion that science routinely observes effect sizes decrease over repeated replications for reasons that are still not well understood'' \citep[p. 579]{schooler2014a}.

Given that underpowered studies are widespread in psychology \citep{cohen1962a,sedlmeier1989a,maxwellPersistenceUnderpoweredStudies2004}, the shrinkage of effect-sizes in replications could be partially explained by the fallacy of ``what does not kill statistical significance makes it stronger'' \citep{loken2017a} and by the trap of the ``winner’s curse'' \citep{button2013a}.

\subsection{The ``What Does Not Kill Statistical Significance Makes It Stronger'' fallacy and the ``Winner's Curse'' trap}

When a statistically significant result is obtained in an underpowered study (e.g., power = 40\%), in spite of the low probability of this event to happen, the result might be seen as even more remarkable. In fact, the researcher might think: ``If obtaining a statistically significant result is such a rare event, and in my experiment I obtained a statistically significant result, it must be a strong one''. This is called the ``what does not kill statistical significance makes it stronger'' fallacy \citep{loken2017a}. The reason why this is a fallacy lies in the fact that it is possible to obtain statistical significance because of many other factors different from the presence of a real effect. Researchers’ degrees of freedom, large measurement errors, and small sample sizes all contribute to create noise in the data, therefore inflating the perhaps true, but small underlying effect. Then, if the procedure used to analyze those data is only focused on a threshold (like in NHST, with the conventional significance level of .05), the noise in the data allows to pass this threshold. 

\begin{figure}[H]

{\centering \includegraphics[width=\maxwidth]{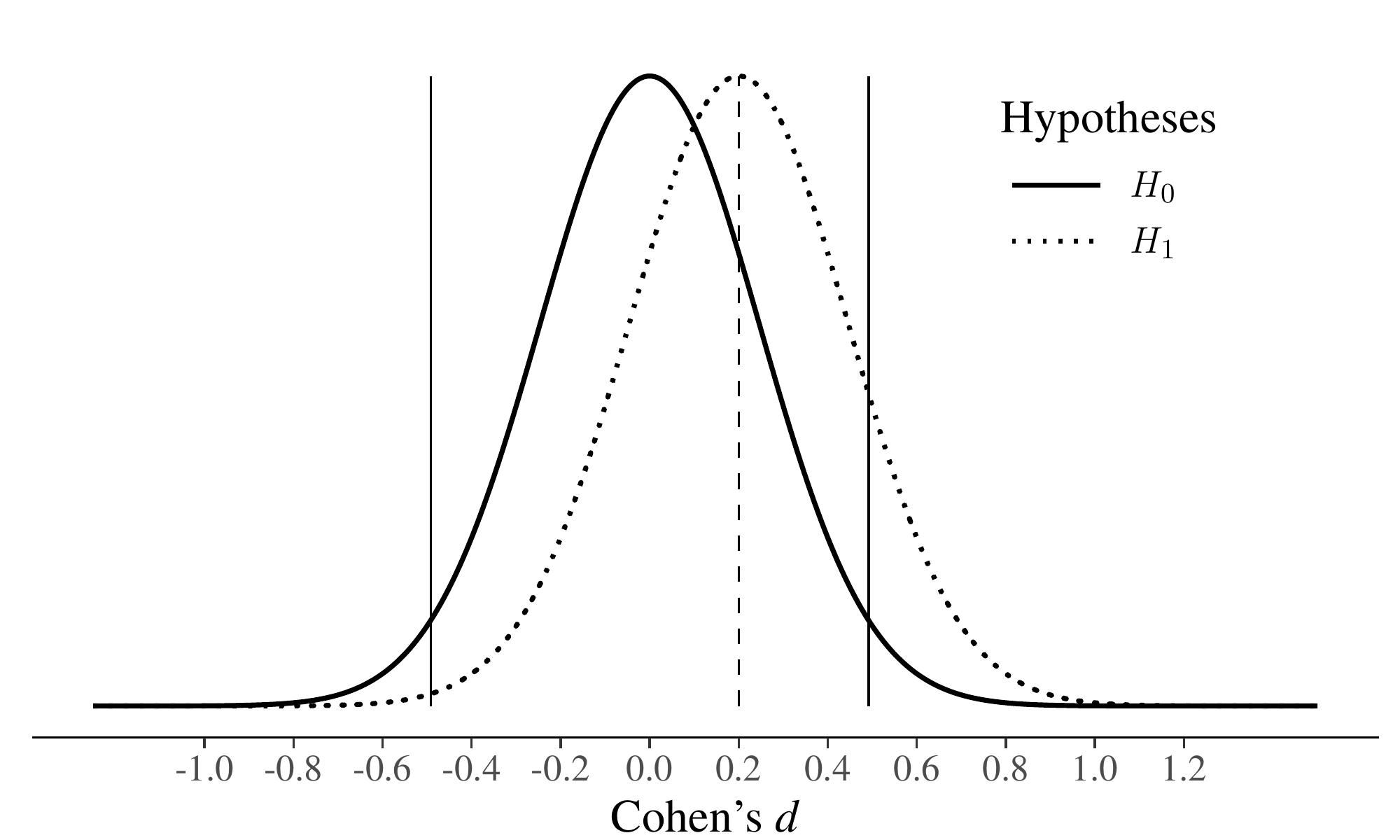} 

}

\caption[Exemplary case study]{The Winner’s Curse. Hypothetical study where the plausible true effect size is small (Cohen’s $d=.20$) and a two tailed independent samples t-test is performed with 33 people per group. In order to reject $H_0$, the researcher has to overestimate the underlying true effect which is indicated by the dashed vertical line. \hspace{\textwidth} Note: the rejection regions of $H_0$, given a significance level of .05, lie outside the vertical black lines} \label{fig:winners.course}

\end{figure}

In these situations, the apparent win in terms of obtaining a statistically significant result is actually a loss, in that the ``the lucky'' scientist who makes a discovery is cursed by finding an inflated estimate of that effect \citep{button2013a}. 
This is called the ``Winner’s curse'', and Figure  \ref{fig:winners.course} shows an example. In this hypothetical situation, the researcher is interested in studying an effect that can plausibly be of small dimensions, e.g. Cohen’s \textit{d} of .20 (see Appendix A, for a detailed description of the calculation and interpretation of Cohen’s $d$). If s/he decides to compare two groups on the outcome variable of interest, using 33 participants per group (and performing a two-tailed test), s/he will never be able to simultaneously reject $H_0$ and find an effect close to what it is plausible in that research field (i.e., .20). In fact, in this underpowered study (i.e., based on a $d$ of .20, the actual power is only 13\%) all the effects falling in the ``rejection regions'' are higher than .49 or smaller than -.49, and .20 falls in the region where the decision rules state that you cannot reject $H_0$ under the NHST approach, or that you can accept $H_0$ under the N-P approach.

\subsection{Beyond Power: The Design Analysis}

As we saw in the previous example, relying solely on the statistical significance of a result can lead to completely misleading conclusions. Indeed, researchers should take into account other relevant information, such as the hypothesized ``plausible effect size'' and the consequent power of the study. Furthermore, to assist researchers with evaluating the results of a study in a more comprehensive way, \citet{gelmanPowerCalculationsAssessing2014a} suggested to consider other two relevant types of errors in addition to the traditional Type I and Type II errors, namely Type $M$ and Type $S$ error (see also, \citealp{gelman2000type,lu2019note}). Specifically, Type $M$ [magnitude] error or \textit{exageration ratio} can be viewed as the the expected average overestimation of an effect that emerges as statistically significant, whereas Type $S$ [sign] error can be viewed as the probability of obtaining a statistically significant result in the opposite direction with respect to the sign of the hypothesized plausible effect size.

Based on this consideration, \citet{gelmanPowerCalculationsAssessing2014a} proposed the term \textit{“design analysis”} to broadly identify the analysis of studies’ properties, such as their statistical power, Type $M$ and Type $S$ error. Moreover, as we shall see in the next paragraph, in design analysis particular emphasis is given on the elicitation and formalization of what can be considered a \textit{plausible effect size} (see also paragraph \ref{wha_doe}) for the study of interest. In this regard, it is important to make a clarification. Although \citet{gelmanPowerCalculationsAssessing2014a} developed design analysis relying on an unstandarized effect size measure (i.e., the difference between two means), in this paper we have adapted their method to deal with Cohen's $d$, a standardized measure of effect size that is more commonly used in psychology (see Appendix A for more details on the reasons that motivated this choice).

Given these premises, the steps to perform design analysis using Cohen's $d$ as a measure of effect size can be summarized in three steps:
\begin{enumerate}
\item A plausible effect size for the study of interest needs to be identified. Rather than focusing on data at hand or on noisy estimates of a single pilot study, the formalization of a plausibile effect size should be based on an extensive theoretical literature review and/or on meta-analyses. Moreover, specific tools (see for example \citealp{o2019expert, zando2019, zondervan2017application}) that allow to incorporate expert knowledge can also be considered to increase the validity of the plausible effect size elicitation process.\footnote{To obtain a more comprehensive picture of the inferential risks associated with their study, we suggest researchers to inspect different scenarios according to different plausible effect sizes and thus to perform more than one design analysis (see for example our application to a real case study in Section \ref{an_ill}).} 
\item Based on the experimental design of the study of interest (in our case, a comparison between two independent groups), a large number of simulations (i.e., 100,000) will be performed according to the identified plausible effect size. This procedure serves to provide information about what to expect if the experiment was replicated an infinite number of times assuming the pre-identified plausible effect as true. 
\item Given a fixed level of Type I error (e.g., .05), power, type $M$ and type $S$ error will be calculated. Specifically, power will be estimated as the ratio between the number of obtained significant results and the number of replicates (i.e., the higher the power, the higher the probability to detect the plausible effect). Type $M$ error will be estimated as the ratio between the mean of the absoulute values of the statistically significant replicated effect sizes and the plausible effect size. In this case, larger values indicate an expected large overestimation of the plausible effect size. Type $S$ error will be the ratio between the number of significant results with opposite sign with regard to the plausible effect size and the the total number of significant results. Put in other terms, type $S$ error estimates the probability of obtaining a significant result in the wrong direction.
\end{enumerate}

Although the procedure may seem complex to implement, at the link \url{https://osf.io/j8gsf/files/} (see also Appendix B) we made available some easy-to-use R functions that allow to perform different types of design analysis for less experienced users. The same functions will also be used in the examples and application presented in this paper. 

To get a first idea of the benefits of design analysis, let us re-analyze the hypothetical study presented in Figure \ref{fig:winners.course}. Specifically, given a plausible effect size equal to $d=.20$ and a sample size of 33 participants per group, a design analysis will highlight the following information: power $= 13$\%, Type $M$ error $= 3.11$, and Type $S$ error $=2$\%.
Despite the low power, which shows that the study has only a 13\% probability to detect the plausible effect size, type $M$ error explicitly indicates that the expected overestimate of a result that will emerge as statistically significant is around 3 times the plausible effect. Furthermore, given a Type $S$ error of 2\%, there is also a non negligible probability of obtaining a significant result in the wrong direction. Overall, the results of design analysis clearly tell the researcher that the study of interest could provide very poor support to both the existence and non-existence of a plausible effect size.

Another advantage of design analysis, which will be better explored in the following sections, is that it can be effectively used in the planning phase of a study, i.e., \textit{prospective design analysis}, as well as in the evaluation of already obtained study results, i.e., \textit{retrospective design analysis}.
For example, in prospective design analysis, considerations concerning power, Type $M$, and Type $S$ error could assist researchers to decide the appropriate sample size for detecting the effect of interest (if it actually exists). In a retrospective design analysis, power, Type $M$ and Type $S$ error (always calculated using the theoretically plausible effect size) can be used to obtain information about the extent to which the results of the study could be exaggerated and/or in the wrong direction. Most importantly, we believe that, engaging in a retrospective design analysis helps researchers to recognize the role of
uncertainty and to make more reasonable statistical claims, especially in those cases at risk of falling in the aformentioned  ``Winner's Curse'' trap.

\begin{figure}[H]
    {\centering \includegraphics[width=\maxwidth]{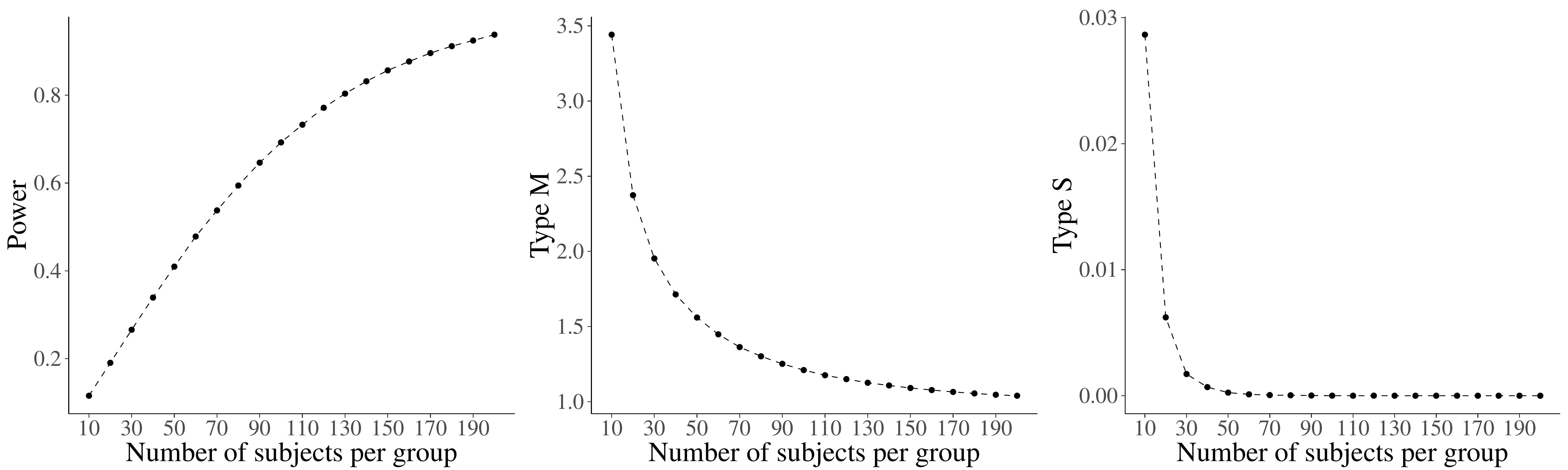}}
    
    \caption[Exemplary case study]{Relationship between sample size and Power, Type $M$ and Type $S$ for a Cohen’s $d$ of .35 in an independent samples t-test. Type I error is set at .05.}\label{fig:pmsn}
\end{figure}

In conclusion, it is important to note that whatever the type of design analysis chosen (prospective or retrospective), the relationships between power, type $M$ error, and type $S$ error are the same. For illustrative purposes, these relationships are graphically displayed  as a function of sample size in Figure \ref{fig:pmsn}. A medium-to-small effect of $d=.35$ (i.e., a reasonable average effect size for a psychological study in the absence of other relevant information, see also Section \ref{an_ill}) was considered as a plausible effect size, and Type I error was set at .05.

As expected, power increases as sample size increases. Moreover, type $M$ and type $S$ error decrease as the size of the sample increases, with the latter showing a much steeper decrease.

From an applied perspective, issues with type $M$ and $S$ errors emerge with underpowered studies which are very common in psychological research. Indeed, as can be seen in Figure \ref{fig:pmsn}, for a power of 40\% (obtained with 48 participants per group), type $M$ error reaches the worrisome value of 1.58; for a power around 10\% (i.e., with 10 participants per group), even type $S$ error becomes relevant (around 3\%).

\subsection{What does ``Plausible Effect Size'' mean?} \label{wha_doe}
 \begin{quotation}
 \begin{flushright}
\textit{``Thinking hard about effect sizes is important for any school of statistical inference\\ {[i.e., Frequentist or Bayesian]}, but sadly a process often neglected.'' \\ \citet[p. 92]{dienesUnderstandingPsychologyScience2008}} 
 \end{flushright}
 \end{quotation}

The main and most difficult point rests on deciding what could be considered a ``plausible effect size''. Although this might seem complex, usually studies are not developed in a void. Hypotheses are derived from theories that, if appropriately formalized in statistical terms, will increase the validity of the inferential process. Furthermore, researchers are commonly interested in knowing the size and direction of effects; as shown above, this corresponds to control for Type $M$ [magnitude] error and type $S$ [sign] error. 

From an epistemological perspective, \citet{kruschke2013a} suggests an interesting distinction between \textit{strong theories} and \textit{weak theories}. Strong theories are those that try to make precise predictions and could be, in principle, more easily disconfirmed. For example, a strong theory could hypothesize a medium-sized positive correlation between two variables. In contrast, weak theories make broader predictions, such as the hypothesis that two variables are correlated without specifying the strength and direction of the correlation \citep{dienesUnderstandingPsychologyScience2008}. The former type allows many more research findings to disconfirm the hypothesis, whereas the latter type allows only the result of no correlation to disconfirm it. Specifically, following Karl Popper (1902–1994), it could be argued that theories explaining virtually everything and being hard to disconfirm risk to be out of the realm of science. Thus, scientific theories should provide at least a hint on the effect that is expected to be observed.

A challenging point is to establish the dimension of this effect. It might seem paradoxical that the researcher has to provide an estimate of the effect size before running the experiment, given that s/he will conduct the study exactly with the aim of finding what that estimate is. However, strong theories should allow to make such predictions, and the way in which science accumulates should provide increasing precision to these predictions.

In practice, it might be undesirable to simply take the estimate found in a pilot study or from a single previous study published in the literature as a ``plausible effect size''. In fact, the plausible effect size refers to what could be approximately the true value of the parameter in the population, whereas the results of pilots or single studies (especially if underpowered) are noisy estimates of that parameter.

In line with \citet{gelmanPowerCalculationsAssessing2014a}, we suggest to use information outside the data at hand, such as literature reviews and/or meta-analyses taking into account issues concernig publication bias \citep{borensteinIntroductionMetaanalysis2009}. Moreover, as stated in the previous paragraph, promising procedures to elicit and formalize expert knowledge should also be considered. It is important to note that, whatever the procedures, all assumptions that will lead to the identification of a plausible effect size must be communicated in a transparent manner, thus increasing the information provided by a study and ensuring more reasonable statistical claims related to the obtained results, whether they are significant or not. 

As we have seen, the identification of a plausible effect size (or a series of plausible effect sizes to explore different scenarios) requires a big effort from the researcher. Indeed, we believe that this kind of reasoning can make a substantial contribution to the planning of robust and replicable studies, as well as to the efficent evaluation of obtained research findings.  

To conclude, we leave the reader with the following question: ``All other conditions being equal, if you had to evaluate two studies of the same phenomenon, the first based on a formalization of the expected plausible effect sizes of interest that is as accurate as possible, and the second one in which the size of the effects of interest was not taken into account, the findings of which study would you believe the most?'' \citep{v2019}.

\section{Prospective and retrospective design analysis}\label{pr_and}
 
To highlight the benefits of design analysis, and to familiarize with the concepts of Type $M$ and Type $S$ errors, we will start from a simple example that is well known in psychological research, i.e., the comparison between the means of two independent groups\footnote{We remind the reader that Appendix B provides a brief explanation of the ad-hoc R functions used in the paper, details on how to reproduce the presented examples and on how to use our R functions for other purposes. The source code of our functions, \texttt{functions\_PRDA.R}, is available at the link \url{https://osf.io/j8gsf/files/}}.

In particular, the goal of our hypothetical case study is to evaluate the differences between two treatments that aim to improve a cognitive ability called $Y$. Both treatments have the same cost, but the first is innovative, whereas the second is traditional. To this end, the researchers recruit a sample of participants who are homogeneous with respect to prespecified relevant study variables (i.e., age, IQ \dots). Next, they randomly assign each participant to one of the two conditions (i.e., innovative vs traditional treatment). After the treatment phase is completed, the means of the two groups are compared.

\subsection{Prospective Design Analysis}

Before collecting data, the researchers decide to plan the appropriate sample size to test their hypotheses, namely that there is a difference between the means of $G1$ (the group to which the innovative treatment is administered) and $G2$ (the group to which the traditional treatment is administered) \textit{vs} there is no difference.

After an extensive literature review concerning studies theoretically comparable to their own, the researchers decide that a first reasonable effect size for the difference between the innovative and the traditional treatment could be considered equal to a Cohen's $d$ of .30 (see Appendix A for a detailed description of the calculation and interpretation of Cohen's $d$).  Due to the possible presence of publication bias \citep{borensteinIntroductionMetaanalysis2009}, which could lead to an overestimation of the effects of published studies, the researchers decide to be more conservative about the estimate of their plausible effect size. Thus, they decide to consider a Cohen's $d$ of .25. Eventually, all researchers agree that a Cohen's $d$ of .25 could also represent a clinically relevant effect in order to support the greater efficacy of the innovative treatment.

Based on the above considerations, the researchers start to plan the sample size for their study. First, they fix the Type I error at .05 and - based on commonly accepted suggestions from the psychological literature - fix the power at .80. Furthermore, to explicitly evaluate the inferential risks connected to their choices they calculate the associated Type $M$ and Type $S$ errors.

Using our \texttt{R} function \texttt{design\_analysis}, they obtain the following results:
\begin{knitrout}
\definecolor{shadecolor}{rgb}{1, 1, 1}\color{fgcolor}\begin{kframe}
\begin{verbatim}
> design_analysis( d=.25 , power=.80 ) 
     d  power      n  typeS  typeM 
  0.25   0.80 252.00   0.00   1.13 
\end{verbatim}
\end{kframe}
\end{knitrout}

Based on the results, to achieve a power of .80, a sample size of 252 for each group is needed (i.e., total sample size = 504). With this sample size, the risk of obtaining a statistically significant result in the wrong direction (Type $S$ error) is practically 0 and the expected exaggeration ratio (Type $M$ error) is  1.13. In other words, the expected overestimation related to effects that will emerge as statistically significant will be around 13\% of the hypothesized plausible effect size.

Although satisfied in terms of expected  type $S$ and type $M$ risks, the researchers are concerned about the economic feasibility of recruiting such a ``large'' number of subjects. After a long discussion, they decide to explore which inferential risks would result for a lower level of power, namely 60\%\footnote{Specifically,  we agree with \citet{gelman2019don} that an 80\% level of power should not be used as an automatic routine, and that requirements of 80\% power could encourage
researchers to exaggerate their effect sizes when planning sample size.}.

Using the function \texttt{design\_analysis}
\begin{knitrout}
\definecolor{shadecolor}{rgb}{1, 1, 1}\color{fgcolor}\begin{kframe}
\begin{verbatim}
> design_analysis( d=.25 , power=.60 ) 
     d  power      n  typeS  typeM 
  0.25   0.60 158.00   0.00   1.30 
\end{verbatim}
\end{kframe}
\end{knitrout}
they discover that: 1) the overall required sample size is considerably smaller (from 504 to 316 = 158 $\times$ 2), thus increasing economic feasability of the study; 2) the Type $S$ error remains negligible (0\%) ; 3) the exaggeration ratio considerably increases (from 1.13 to 1.30); thus, an effect that will emerge as statistically significant will be on average 130\% of the hypothesized plausible effect size. 

The researchers now need to make a decision. Even though, from a merely statistical point of view, the optimal choice would be to consider a power of 80\%, other relevant aspects must be evaluated, such as the possibility to obtain additional funding, the practical implications of an expected overestimation of the plausible effect size, and the phase of the study (i.e., preliminary/exploratory, intermediate or final/confirmatory).

Whatever the decision, the researchers have to be aware of the inferential risks related to their choice. Moreover, when presenting the results, they have to be transparent and clear in communicating such risks, thus highlighting the uncertainty associated with their conclusions.

\subsection{Retrospective Design Analysis}

To illustrate the usefulness of retrospective design analysis, we refer to the example presented in the previous paragraph. However, we introduce three new scenarios which can be considered as representative of what commonly occurs during the research process:
\begin{itemize}
\item \textbf{Scenario 1 (S1): Evaluating sample size based on a single published study.}\footnote{Even though in this paper we strongly recommend not to plan sample size based on a single study, we propose this example to further emphasize the inferential risks associated with the information provided by a single underpowered study.}\\
Imagine that the researchers decide to plan their sample size based on a single published study in the phase of formalizing a plausible effect size, either because the published study presents relevant similarities with their own study, or because there are no other published studies available.\\
\textit{Question}: What type of inferential risks can be associated with this decision?\\
\textit{Issues}: Using a single study as a reference point without considering other sources (e.g., theoretical framework, experts’ opinion, or a meta-analysis), especially when the study has a low sample size and/or the effect of interest is small, can lead to use an excessively optimistic estimate of the effect size to plan an appropriate sample size \citep{gelmanPowerCalculationsAssessing2014a}.
\item \textbf{Scenario 2 (S2): Difficulty in recruiting the planned number of research participants.}\\
Imagine that, due to unforeseen difficulties (e.g., insufficient funding), the researchers are not able to recruit the pre-planned number of participants as defined based on prospective design analysis.\\
\textit{Question}: How to evaluate the inferential risks associated with the new reduced sample size? How to communicate the obtained results?\\
\textit{Issues}:  Researchers are often tempted to evaluate the results of their study based on the observed effect size. This procedure, known as ``post-hoc power analysis'', has been strongly criticized and many statistical papers explicity advise against its use (see for example, \citealp{gelman2019don, goodman1994use}). Indeed, to evaluate the information provided by the obtained results, researchers should use the a piori plausible effect size, i.e., the one formalized before collecting their data. 
\item \textbf{Scenario 3 (S3): No prospective design analysis because the number of participants is constrained.}\\
Imagine the number of participants involved in the study have specific characteristics which make it impossible to yield a large sample size, or that the type of treatment is particularly expensive and therefore it cannot be tested on a large sample. In this case, the only possibility is to recruit the largest possible number of participants.\\
\textit{Question}: What level of scientific quality can be provided by the results?\\
\textit{Issues}:  Although study’s results can provide a useful contribution to the field, there are several associated inferential risks that the researchers need to communicate in a transparent and constructive way.
\end{itemize}

As we will see below, retrospective design analysis can be a useful tool to deal with the questions and the issues raised across all three scenarios.

For the sake of simplicity and without loss of generalizability, suppose that in each of the three scenarios the researchers obtained the same results (see Table \ref{tab:mean_comp}). \\

\begin{table}[h]
\centering
\caption{Comparison of the cognitive skill $Y$ between the two groups}
\label{tab:mean_comp}
\begin{tabular}{llccccc}
\hline
Group & $n$ & $M$ & $SD$ & $t(df)$ & $p$ & Cohen's $d$ ($95\% \, CI$) \\ \hline
Innovative treatment  & 31 & 114 & 16 & 3.496 (60) & 0.001 & .90 (0.38-1.43) \\
Traditional treatment & 31 & 100 & 15 &  &  & \\ \hline
\end{tabular}
\end{table}

At a first glance, the results indicate a statistically significant difference in favor of the innovative treatment (see Table \ref{tab:mean_comp}), with a large effect size (i.e., $d=0.90$). However, the 95\% confidence interval for Cohen's $d$ is extremely wide, suggesting that both medium-small (i.e., $d=0.38$) and very large (i.e., $d=1.43$) effects are consistent with the observed data.

A closer look indicates that the estimated effect size seems too large when compared with the initial guess of the researchers (i.e. $d=.25$). Furthermore,
an estimated $d$ of 0.90 seems, in general, implausibly large for a difference between two cognitive treatments (see also Appendix A). The latter interpretation seems to be also supported by the fact that the hypothesized plausible effect size is not even included in the estimated confidence interval.
Overall, in order to prevent the aforementioned ``Winner’s Curse'' and ``What Does Not Kill Statistical Significance Makes It Stronger'' heuristics, results must be evaluated and eventually communicated with caution and skepticism. 

To obtain a clearer picture of the inferential risks associated with the observed results, we can perform retrospective design analysis using $d=.25$ as plausible effect size and 31 participants per group as sample size:
\begin{knitrout}
\definecolor{shadecolor}{rgb}{1, 1, 1}\color{fgcolor}\begin{kframe}
\begin{verbatim}
> design_analysis( n=31, d=.25 ) 
power typeS typeM 
 0.16  0.01  2.59 
\end{verbatim}
\end{kframe}
\end{knitrout}

As can be seen, the power is markedly low (i.e, only 16\%) and the Type $M$ error even suggests an expected overestimation around two and a half times the plausible effect size. Lastly, the Type $S$ error, although small, indicates a 1\% risk of obtaining a significant result in the wrong direction (i.e., the traditional treatment is better than the innovative treatment). Let's see how this information could be helpful to deal with the three presented scenarios.

In S1, the researchers took a single noisy estimate as the plausible effect size from a study that found a ``big'' effect size (e.g., 0.90). The retrospective design analysis shows what happens if the plausible effect size is, in reality, much smaller (i.e., 0.25). Specifically, given the low power and the high level of Type $M$ error, researchers should definitely abandon the idea of planning their sample size based on a single published study. Furthermore, issues regarding the presence of Questionable Research Practices \citep{arrison2014responsible, john2012measuring}  and Questionable Measurement Practices \citep{flake2019measurement} in the considered published study must be at least explored. From an applied perspective, researchers should continue with a more comprehensive literature review  and/or consider the opportunity to use an effect size elicitation procedure based on experts' knowledge \citep{o2019expert, zondervan2017application}.\\

In S2, to check the robustness of their results, researchers might initially be tempted to conduct a power analysis based on their observed effect size ($d=0.90$). Acting in this way, they would obtain a completely misleading post-hoc power of 94\%. In contrast, the results of retrospective design analysis based on the a-priori plausible effect size ($d=.25$) highlight the high level of inferential risks related to the observed results. From an applicative perspective, researchers should be very skeptical about their observed results. A first option could be to replicate the study on an independent sample, perhaps asking for help from other colleagues in the field. In this case, the effort to recruit a larger sample could be well-justified based on the retrospective design analysis.

In S3, given the low power and the high level of Type $M$ error, results should be presented as merely descriptive by clearly explaining the uncertainty that characterizes them. Researchers should first reflect on the possibility of introducing improvements to the study protocol (i.e., improving the reliability of the study variables). As a last option, if improvements are not considered feasible, the researchers might consider not continuing their study. 

Despite its advantages, we need to emphasize that design analysis should not be used as an \textit{automatic problem solver machine}: ``Let's pull out an effect size \dots let me see the correct sample size for my experiment''. In other words, to obtain reliable scientific conclusions there is no ``free lunch''. Rather, psychologists and statisticians have to work together, case by case, to obtain a reasonable effect size formalization and to evaluate the associated inferential risks. Furthermore, researchers are encouraged to explore different scenarios via sensitivity analysis (see, Section \ref{an_ill}) to better justify and optimize their choices.

\section{Incorporating uncertainty concerning effect size formalization in retrospective design analysis}\label{inc_unc}

As shown in the previous examples, a key point both in planning (i.e., prospective design analysis) and in evaluating (i.e., retrospective design analysis) a study is the formalization of a plausible effect size. Using a single value to summarize all external information and previous knowledge with respect to the study of interest can be considered an excessive simplification. Indeed, all uncertainty concerning the magnitude of the plausible effect size is not explicitly taken into consideration. In particular, the level of heterogeneity emerging from the examination of published results and/or from different opinions of the consulted experts is poorly formalized. The aim of this paragraph is to propose a method that can assist researchers to deal with these relevant issues. Specifically, we will focus on the evaluation of the results of a study (i.e., retroprospective design analysis).

Our method could be summarized in the following three steps: 1) defining a lower and an upper bound within which the plausible effect size can reasonably vary; 2) formalizing an appropriate probability distribution that reflects how the effect size is expected to vary; 3) conducting the associated analysis of Power, Type $M$ and Type $S$ error.

To illustrate the procedure, we take the study presented in Table \ref{tab:mean_comp} as reference. Let us now hypothesize that after a thorough evaluation of external sources, the researchers conclude that a plausible effect size could reasonably vary between .20 and .60 (instead of specifying a too simplistic single point value). It should be noted that, from a methodological perspective, the specification of a ``plausible interval'' can be considered an efficient and informative starting point to elicit the researchers' beliefs \citep{o2019expert}. 

\begin{figure}[H]
    {\centering \includegraphics[width=\maxwidth]{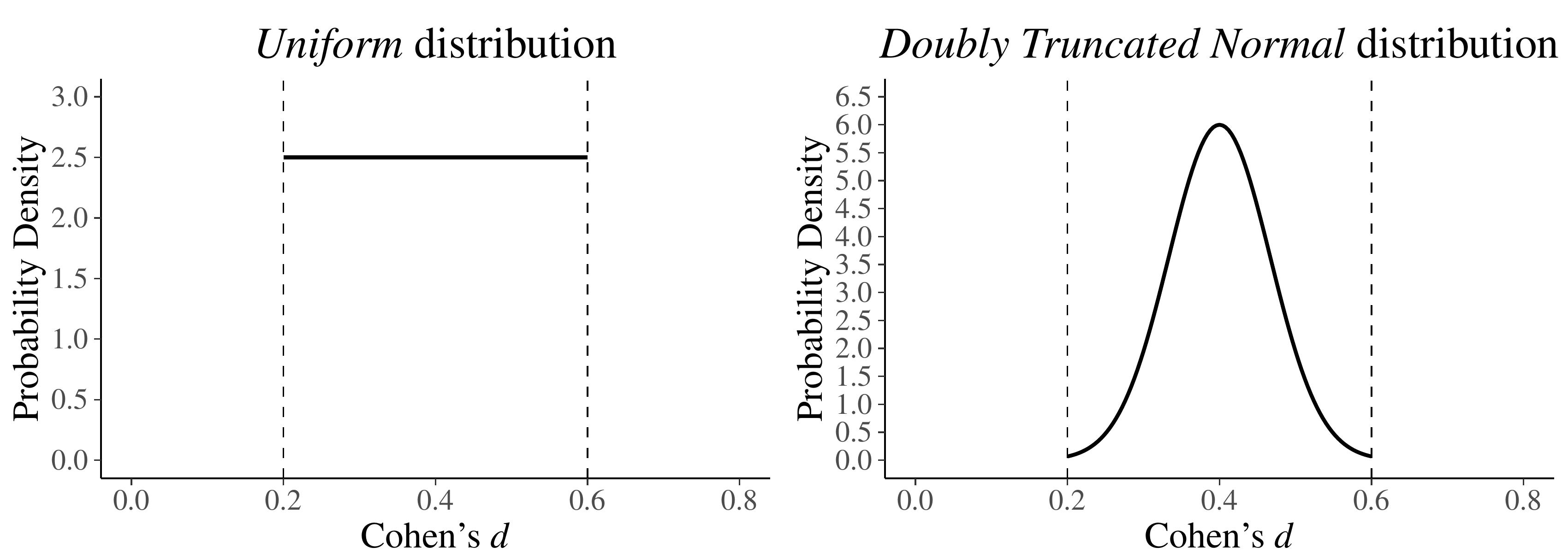}}
    
    \caption[Exemplary case study]{Different ways to formalize a plausible interval for the effect size $d$. In the left panel, a Uniform distribution with lower bound = .20 and upper bound = .60 is used. In the right panel, a doubly truncated Normal distribution with lower bound = .20, upper bound = .60, mean = .40 and standard deviation equal to $\frac{1}{6}$  the lenght of the interval (i.e., $\frac{.60-.20}{6}=.067$) is used.} \label{fig:dintervals}

\end{figure}

At this point, a first option could be to assume that, within the specified interval, all effect size values have the same probability of being true. This assumption can be easily formalized using a Uniform distribution as the one shown in Figure \ref{fig:dintervals} (left panel). 

However, from an applied point of view  it is rare for the researcher to expect that all values within the specified interval have the same plausibility. Indeed, in general conditions, it is more reasonable to believe that values around the center of the interval (i.e., .40 in our case) are more plausible, and that their plausibility gradually decreases as they move away from the center. This expectation can be directly formalized in statistical terms  using the so-called ``doubly truncated Normal distribution''. On an intuitive level (for a more complete description see \citealp*{burkardt2014truncated}), the doubly truncated Normal distribution can be seen as a Normal distribution whose values are forced to vary within a specific closed interval. In case of the formalization of the plausible effect size, we propose to use a doubly truncated Normal distribution with the following parameters: a lower and an upper bound according to the prespecified plausible interval, a mean fixed at the center of the interval, and a standard deviation that reflects the hypothesized uncertainty around the center.  A standard deviation of $\frac{1}{10}$ the length of the chosen interval will produce a substantially Normal distribution. Higher values, like $\frac{1}{6}$ the length of the interval (see right panel of Figure \ref{fig:dintervals}) will lead to normal-like distributions with increased probability on the tails, thus reflecting greater uncertainty around the center.

Coming back to our example, suppose that the researchers want to evaluate the study of interest assuming a plausible interval for Cohen’s  $d$ as the one represented in Figure \ref{fig:dintervals}.
Using the ad-hoc function \texttt{design\_est}\footnote{The idea behind this function is simple. First, we sample a large number (e.g., 100,000) of effect sizes $d$ from the probability distribution associated with the plausible interval. Then, for each $d$ we calculate power, type $M$ and type $S$ error based on the sample size of the two groups involved in the comparison, and by considering the centre of the plausible interval as the most plausible effect size. In this way, a distribution for each of the three indices is finally obtained. In the output of the function, the means of the three distributions are presented as a summary value. For additional details, see Appendix B, which also shows (in section `` design\_est'') how to obtain the expected distribution of power, Type $M$ and Type $S$ errors, given the plausible interval for $d$.} they will obtain the following information \color{black}:

\begin{knitrout}\small
\definecolor{shadecolor}{rgb}{1, 1, 1}\color{fgcolor}\begin{kframe}
\begin{verbatim}
> design_est(n1=31, n2=31, target_d_limits=c(.20,.60), distribution="normal") 
power typeS typeM 
 0.35  0.00  1.73 
\end{verbatim}
\end{kframe}
\end{knitrout}
To summarize, this information suggests that the results of the study of interest  (see, Table \ref{tab:mean_comp}) should be taken very cautiously. Indeed, the expected power is low (35\%), and the expected overestimation of the most plausible effect size (i.e., $d = .40$) is around 73\%. Furthermore, it is important to note that the observed effect size of .90 falls abundantly outside the pre-specified plausible interval of .20 - .60, thus supporting the idea that the study of interest is clearly overestimating the actual magnitude of the effect.

In general, when the observed effect size falls outside the prespecified plausible interval, we can conclude that the observed study is not coherent with our theoretical expectations. On the other hand, we could also consider that our plausible interval may be unrealistic and/or poorly formalized.  
In these situations, researchers should be transparent and propose possible explanations that could be very helpful in understanding the phenomenon under study.  Although this way of reasoning requires a notable effort, the information provided will lead to a more comprehensive inference than the one deriving from a simplistic dichotomous decision  (i.e., ``reject / do not reject'') typical of the NHST approach. Indeed, in this approach the hypotheses are poorly formalized,  and power, Type $M$ and Type $S$ errors are not even considered.

\section{An illustrative application to a case study} \label{an_ill}

To illustrate how design analysis could enhance inference in psychological research, we will consider a real case study. Specifically, we will focus on Study 2 of the published paper ``A functional basis for structure-seeking: Exposure to structure promotes willingness to engage in motivated action'' \citep{kay2014functional}. 

The paper presents 5 studies arising from findings showing that human beings have a natural tendency to perceive structure in the surrounding world. Various social psychology theories propose plausible explanations which share a similar assumption that had never been tested before, that is, perceiving a structured world could increase people’s willingness to make efforts and sacrifices towards their own goals. In Study 2, the authors decided to test this hypothesis by randomly assigning participants to two different conditions differing in the type of text they had to read. In the ``random'' condition, the text conveyed the idea that natural phenomena are unpredictable and random, whereas in the ``structure'' condition the phenomena were described as predictable and systematic. The outcome measure was the willingness to work towards a goal that each participant chose as their ``most important''. The expected result was that participants in the  ``structure'' condition would report a higher score in the measure of goal directed behavior than those in the ``random'' condition.

\vspace{1em}

\textbullet\ \textbf{Prospective design analysis}

As we saw in the previous paragraphs, before collecting data is fundamental to plan an appropriate sample size via prospective design analysis. In this case, given the relative novelty of Study 2, is hard to identify a single plausible value for the size of the effect of interest. Rather, it seems more reasonable to explore different scenarios according to different plausible effect sizes and power levels. We will start with a minimum  $d$ of .20,  so that the study is planned to detect at least a ``small'' effect size. If the final results do not reach statistical significance, the researchers could conclude that it is unlikely that the true effect is equal to or greater than .20, and eventually decide whether it is worth replicating the study, perhaps by modifying their protocol.

As the most plausible effect size, we will consider $d=.35$, which could be considered – at least in our opinion - a typical average level to test a hypothesis in psychological research in the absence of informative external sources (see for example the results reported in \citealp{opensciencecollaborationEstimatingReproducibilityPsychological2015a})\footnote{In the \citet{opensciencecollaborationEstimatingReproducibilityPsychological2015a}, the authors conducted replications of 100 experimental and correlational studies published in three psychology journals using high powered designs and original materials when possible. They found an average effect size of $r = .197$, i.e. $d = .41$. Given the heterogeneity of the 100 studies, we propose to use a more conservative value as being representative of a typical average effect in psychology. Overall, it should be noted that all the pre-specified values of $d$, albeit plausible, are not based on a thorough theoretical revision and/or on the formalized knowledge of experts in the field. Indeed, an appropriate use of the latter two external sources would undoubtedly contribute to produce more reliable results, but discussion of these strategies is beyond the scope of this paper.}.
As extrema ratio, we will include also a d of .5, which in the words of Jacob Cohen can be referred to as ``differences that are large enough to be visible to the naked eye'' (see \citealt[p.26]{cohenStatisticalPowerAnalysis1988} and Appendix A), and that, given the experiment under investigation, could be viewed as an extremely optimistic guess. Finally, to take issues concerning the feasibility of the study into account, we will also consider two levels of power, namely 80\% and 60\%.

Overall, our ``sensitivity'' prospective design analysis (see Table \ref{tab:casestudy1}) suggests that the sample size chosen by the authors ($n=67$) is inadequate. Indeed, even in the least reasonable scenario ($d$ = .50, power = .60), a minimum of 80 participants is required. Furthermore, is should be noted, that the associated Type $M$ error is considerably high, i.e. 130\%, signaling a high risk of overstimating the plausible effect.

\begin{table}[h]
\centering
\caption{Sample size, Type M and Type S error by power and plausible effect size. Type I error is fixed at .05}
\label{tab:casestudy1}
\begin{tabular}{cccccc}
\hline
Power                & Cohen's $d$& $n$ (per sample)   & Total $n$          & Type $M$ error        & Type $S$ error    \\
\hline
\multirow{3}{*}{.80} & 0.20 & 392 & 784 &                     &                    \\
                     & 0.35 & 130 & 260 & 1.13 & 0.00\\       
                     &0.50  & 64 & 128 &                     &                    \\ 
\hline                   
\multirow{3}{*}{.60} &0.20  & 244 & 488 &                     &                    \\
                     &0.35  & 82 & 164 & 1.30 & 0.00\\ 
                     &0.50  & 40 & 80 &                     &          \\ 
\hline                    
\end{tabular}
\end{table}
A good compromise could be to consider the second scenario ($d$ = .35, power = .80), which requires a total sample size of 260, guaranteeing optimal control of the Type $M$ error. After conducting the study with this sample size, a significant result would lead to accept the researcher's hypothesis, while a non-significant result would indicate that if an effect exists, it will presumably be less than .35. Whatever the result, the researchers could eventually present their findings in a transparent and informative way. In any case, the results could be used to improve scientific progress. As an example, other researchers could fruitfully use the observed results as a starting point for a replication study.

\textbullet\ \textbf{Retrospective design analysis}

Let us now evaluate Study 2 from a retrospective point of view. Based on their results ($M_\text{structure} = 5.26$, $SD_\text{structure}=0.88$, $M_\text{random} = 4.72$, $SD_\text{random}= 1.32$, $n_\text{total}=67$; $t(65)=2.00$, $p=.05$, $\text{Cohen’s}\,d = 0.50$)\footnote{The authors reported only the total sample size ($n=67$). Since participants were randomly assigned to each of the two experimental conditions, in the following we will assume, without loss of generalizability, that 34 participants were assigned to the ``structure'' condition, and 33 to the ``random'' condition.}, the authors  concluded that ``participants in the structure condition reported higher willingness to expend effort and make sacrifices to pursue their goal compared to participants in the random condition.'' \citet[p. 487]{kay2014functional}, thus supporting their initial  hypothesis.

To evaluate the inferential risks associated with this conclusion, we now run a sensitivity retrospective design analysis on the pre-identified plausible effect sizes  (i.e., $d = .20$, $d = .35$, $d = .50$).

In line with the results emerging from prospective analysis, the retrospective design analysis indicates that the sample size used in Study 2 exposes to high inferential risks. In fact, both for a plausible effect of $d=.20$ (power = .13, type $M =$ 3.06, type $S=$ 2\%) and for a plausible effect of $d=.35$ (power = .29, type $M =$ 1.86, type $S=$ 0\%), the power is very low and the Type $M$ error reaches worrying levels. For a $d$ of .50 (chosen on the basis of plausible effects and not based on the results observed in Study 2), the Type $M$ error is 1.40, indicating an expected overestimate of 40\%. Furthermore, the power is .52, suggesting that if we replicate the study on a new sample with the same number of participants, the probability of obtaining a significant result will be around the chance level.

We also evaluate the results of Study 2 by performing a retrospective design analysis using the method presented in Section \ref{inc_unc}. Specifically, we use a doubly truncated normal distribution centered at .35 (i.e., the most plausible effect size) with a plausible interval of .25 - 45. As could be expected, the results (i.e., power = .29, type $M =$ 1.86, type $S=$ 0\%) substantially confirm what emerged from sensitivity retrospective design analysis.

In summary, our retrospective design analysis indicates that, although statistically significants the results of Study 2 are inadequate to support the authors' conclusions.

As mentioned at the beginning of this paragraph, the Study 2 of \citet{kay2014functional} was selected for illustrative purposes and in a constructive perspective. For a more comprehensive picture, we invite interested readers to consult the ``Many Labs 2 project'' \citep{klein2018many}, which showed that with a large sample size ($n=6506$) the original conclusion of Study 2 cannot be supported (i.e., $t(6498.63) = -0.94$, $p = .35$, $d = -0.02$, $95\%CI = [-0.07, 0.03]$, as well as the subsequent response of the original authors \citep{laurin2018structure}.

\section{Discussion and Conclusions}\label{dis_and}

In psychological research, statistical inference is often viewed as an isolated procedure which limits itself to the analysis of data that have already been collected. In this paper, we argue that statistical reasoning is necessary both at the planning stage and when interpreting the results of a research project. To illustrate this concept, we built on and further developed Gelman and Carlin’s (2014) idea of ``prospective and retrospective design analysis''.

In line with recent recommendations \citep{cumming2014new}, design analysis involves an in-depth reasoning on what could be considered as a plausible effect size within the study of interest. Specifically, rather than focusing on a single pilot or published study, we underlined the importance of using information outside the data at hand, such as extensive literature reviews and meta-analytic studies taking issues related to publication bias into account. Furthermore, we introduced the potentials of elicitation of expert knowledge procedures (see for example \citealp{o2019expert, zondervan2017application}). Even though these procedures are still underutilized in psychology, they could provide a relevant contribution to the formalization of research hypotheses.

Moving beyond the simplistic and often misleading distinction between significant and non significant results, design analysis allows researchers to quantify, consider, and explicitly communicate two relevant risks associated with their inference, namely exaggeration ratio (Type $M$ error) and sign error (Type $S$ error). As illustrated in the paper, the evaluation of these risks is particularly relevant in studies which investigate small effect sizes in the presence of high levels of intra- and interindividual variability, with a limited sample size – a situation that is quite common in psychological research.

Another important aspect of design analysis is that it can be usefully carried out both in the planning phase of a study (i.e., prospective design analysis) and to evaluate studies that have already been conducted (i.e., retrospective design analysis), reminding researchers that the process of statistical inference should start before data collection and does not end when the results are obtained. In addition, design analysis contributes to have a more comprehensive and informative picture of the research findings through the exploration of different scenarios according to different plausible formalizations of the effect of interests.

To familiarize the reader with the concept of design analysis, we included several examples and an application to a real case study. Furthermore, in addition to the classic formalization of the effect size with a single value, we proposed an innovative method to formalize uncertainty and previous knowledge concerning the magnitude of the effect via probability distributions within a Frequentist framework. Although not directly presented in the paper, it is important to note that this method could also be efficiently used to explore different scenarios according to different plausible probability distributions.

Finally, to allow researchers to use all the illustrated methods with their own data, we also provided two easy-to-use \texttt{R} functions (see also Appendix B) which are available at the Open Science Framework (OSF) at the link \url{https://osf.io/j8gsf/files/}.

For the sake of simplicity, in this paper we limited our consideration to Cohen’s $d$ as an effect size measure within a Frequentist approach. However, the concept of design analysis could be extended to more complex cases and to other statistical approaches. For example, our \texttt{R} functions could be directly adapted to other effect size measures, such as Hedges' $g$, Odds Ratio, $\eta^2$ and $R^2$. Moreover, concerning the proposed method to formalize uncertainty and prior knowledge, other probability distributions beyond those proposed in this paper (i.e., the uniform and the doubly truncated normal) could be easily added. This was one of the main reasons behind the choice to use resampling methods to estimate power, Type $M$ and Type $S$ errors instead of using an analytical approach.  

Also, it is important to note that our considerations regarding design analysis could be fruitfully extended to the increasingly used Bayesian methods. Indeed, our proposed method to formalize uncertainty via probability distributions finds its natural extension in the concept of Bayesian prior.
Specifically, design analysis could be useful to evaluate the properties and highlight the inferential risks (such as type $M$ and type $S$ errors) associated with the use of Bayes Factors and parameter estimation with credible Bayesian intervals.

In sum, even though a design analysis requires big effort, we believe that it has the potential to contribute to planning more robust studies and promoting better interpretation of research findings. More generally, design analysis and its associated way of reasoning helps researchers to keep in mind the inspiring quote presented at the begininng of this paper regarding the use of statistical inference: ``Remember ATOM''.

\section*{Conflict of Interest Statement}
 
The authors declare that the research was conducted in the absence of any commercial or financial relationships that could be construed as a potential conflict of interest.

\section*{Author Contributions}

GA conceived the original idea and drafted the paper; GB, CZC and ET contributed to the development of the original idea and drafted sections of the manuscript ; MP and GA wrote the R functions; GA, MP and CZC took care of the statistical analysis and of the graphical representations. LF and AC provided critical and useful feedback. All authors contributed to manuscript revision, read and approved the submitted version.

\bibliography{bibliography}

\newpage
\section*{Appendix A: Cohen's d}

\subsection*{Cohen's $\bm{d}$ definition}

When considering the difference between two groups, the raw mean difference has an intuitive and meaningful interpretation if the outcomes are measured using a widespread scale (i.e., meters for height or kilos for weight). This allows us to easily share and compare the results of different studies. However, in the behavioural and social sciences, the same outcomes are often evaluated with different instruments, each of which has its own scale (i.e., questionnaire or test scores). Thus, it is difficult to interpret and compare study results when the outcomes are assessed with different measurement scales.

To overcome this issue, \citet[p.21]{cohenStatisticalPowerAnalysis1988} proposed what he described as “a pure number [...], freed of dependence upon any specific unit of measurement”, the now-famous Cohen's $d$. Under the assumptions of normality and homogeneity of variance, Cohen's d ($\delta$) is defined as the raw difference between two population means ($\mu_A$ and $\mu_B$) divided by the common standard deviation ($\sigma$):
\begin{equation}
  \delta = \frac{\mu_A-\mu_B}{\sigma}
\end{equation}

Cohen's $d$ is a standardized measure of effect size that allows to express differences in terms of the variability of the phenomena of interest irrespective of the original measurement unit. It is a useful solution when researchers utilize raw units which are quite arbitrary or lack meaning outside their investigation \citep{cohenStatisticalPowerAnalysis1988}.  A Cohen's $d$ of 0.1 means that the difference between the two population means is one-tenth of the common standard deviation.

\citet{borensteinIntroductionMetaanalysis2009} underline the importance of distinguishing between $\delta$, the population Cohen's $d$ value, and $d$, the estimated Cohen's $d$ value from the sampled groups given by:
\begin{equation}
  d = \frac{\bar{X}_A-\bar{X}_B}{S_{pooled}}
\end{equation}

In the numerator, $\bar{X}_A$ and $\bar{X}_B$ are the sample means in the two groups. In the denominator, $S_{pooled}$ is the pooled standard deviation:

\begin{equation}
  S_{pooled} = \sqrt{\frac{(n_A-1)S^2_A-(n_B-1)S^2_B}{n_A+n_B-2}}
\end{equation}
where $n_A$ and $n_B$ are the two sample sizes, and $S^2_A$ and $S^2_B$ are the standard deviations in the two groups.

\subsection*{Cohen's $\bm{d}$ interpretation}

In well-established areas of study, the definition of a relevant effect size in terms of Cohen's $d$ offers no particular difficulty. Population $\sigma$ is normally already known or easy to estimate, and differences of interest are easily defined from the research context. Thus, researchers may already know which is the effect size of interest to evaluate, for example, the effectiveness of a specific treatment. On the contrary, in the case of less known areas, or when newly developed measures are employed, the definition of an effect size of interest may not be so simple.

In these cases, \citet{cohenStatisticalPowerAnalysis1988} proposed some conventional operational definitions to interpret effect sizes. He suggested indicative values of $d$ for ``small'', ``medium'', and ``large'' effect sizes.

\begin{itemize}
  \item{\textit{Small effect size: d = .2}; This refers to small differences that are difficult to detect, such as  approximately the size of the difference in mean height between 15- and 16-year-old girls.} 
  \item{\textit{Medium effect size: d = .5}; This refers to differences that are ``large enough to be visible to the naked eye'' (p.26). For example, the magnitude of the difference in height between 14- and 18-year-old girls.}
  \item{\textit{Large effect size: d = .8}; This refers to very obvious differences, such as the mean difference in height between 13- and 18- year-old girls.}
\end{itemize}

\begin{figure}[!h]

{\centering \includegraphics[width=\maxwidth]{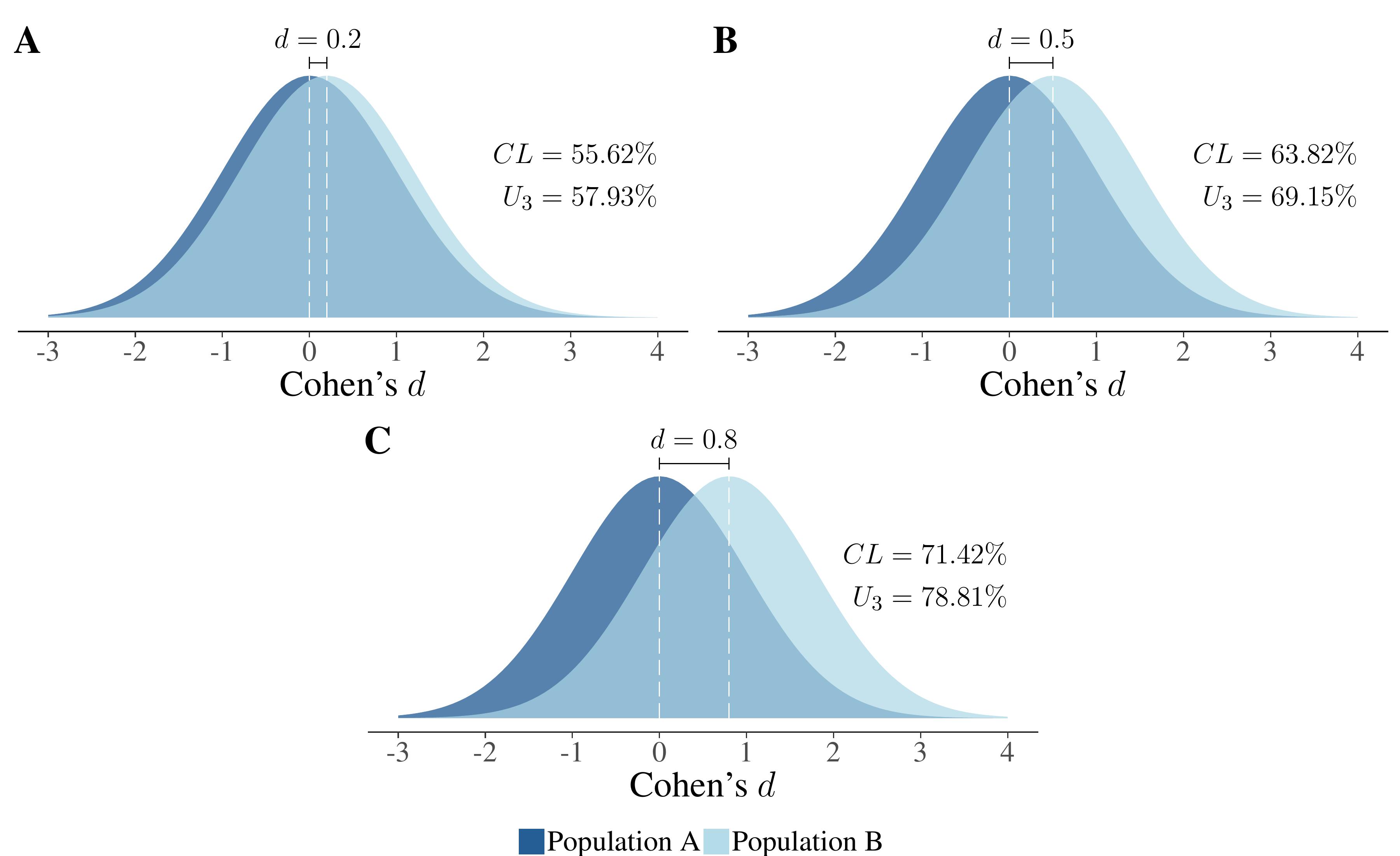} 
}

\caption[Exemplary case study]{$CL$ and $U_3$ values according to effect sizes. (\textbf{A}) In the case of a small effect ($d=.2$), there is a 56\% probability that a random subject from population B has a higher score than a random subject from population A ($CL=.56$) and 58\% of population B is above the mean of population A ($U_3=.58$). (\textbf{B}) In the case of a medium effect ($d=.5$), there is a 64\% probability that a random subject from the population B has a higher score than a random subject from population A ($CL=.64$) and 69\% of population B is above the mean of the population A ($U_3=.69$). (\textbf{C}) In case of a large effect ($d=.8$), there is a 71\% probability that a random subject from population B has a higher score than a random subject from population A ($CL=.71$) and 79\% of population B is above the mean of population A ($U_3=.79$).}\label{fig:effects}

\end{figure}

Another way to interpret and make sense of Cohen's $d$ values is to consider the Common Language effect size statistic \citep[CL;][]{ruscioProbabilitybasedMeasureEffect2008}, or Cohen's measure of non-overlap $U_3$ \citep{cohenStatisticalPowerAnalysis1988}. The former is defined as the probability that a randomly chosen member of population B scores higher than a randomly chosen member of population A. The latter is defined as the percentage of the population B which exceeds the mean of the population A. Figure \ref{fig:effects} shows $CL$  and $U_3$ values for small, medium, and large effect sizes.

However, as suggested by \citet[p.25]{cohenStatisticalPowerAnalysis1988}, “The terms \textit{small}, \textit{medium}, and \textit{large} are relative, not only to each other but to the area of behavioural science or even more particularly to the specific content and research method being employed in any given investigation”. These values are only conventional references in the absence of any other information. Researchers should aim to define their own criteria, according to their specific research objectives and the related
costs-benefits ratio. In some fields, even small changes could result in valuable gains. 

Finally, it is important to underline an aspect that is often neglected when dealing with Cohen's $d$. It should be remembered that Cohen's $d$ depends on the pooled standard deviation (i.e., increasing levels of standard deviation are associated with lower values of Cohen's $d$). Given that the pooled standard deviation partly reflects the accuracy of the measure used in a study, in the planning phase researchers should select measures that are as accurate as possible. Furthermore, when evaluating effect sizes of other studies, considerations about the accuracy of the utilized measure(s) should always be taken into account.

\clearpage
\section*{Appendix B: R functions for design analysis}
\subsection*{Preliminary notes}
The \texttt{R} functions presented in the paper to perform design analysis are:
\begin{itemize}
\item \texttt{design\_analysis( )}
\item \texttt{design\_est( )}
\end{itemize}
These functions are detailed described in the following section, and their code (i.e., \texttt{PRDA.R}) is available at the Open Science Framework (OSF) at the link \url{https://osf.io/j8gsf/files/}.

In the last section of this Appendix, all examples in the paper are also reproduced using the aforementioned functions. It should be noted that results might slightly differ because both functions follow a simulation approach. To obtain more stable results, it is possible to increase the default number of iterations.

Readers can use the functions to easily perform prospective and retrospective design analysis on their own data. 

Furthermore, the \texttt{R} code can be used as a starting point to extend design analysis to more complex cases than the one presented (i.e., the differences between two independent groups considering Cohen’s $d$ as an effect-size measure) and that were beyond the scope of this paper.

\newpage
\subsection*{R functions}
To use our \texttt{R} functions, first download the file \texttt{PRDA.R} at the link \url{https://osf.io/j8gsf/files/}. To load the functions, simply type:
\begin{knitrout}\small
\definecolor{shadecolor}{rgb}{0.988, 0.988, 0.988}\color{fgcolor}\begin{kframe}
\begin{verbatim}
> source("PRDA.R")
\end{verbatim}
\end{kframe}
\end{knitrout}
\textit{Note: For the correct use of the functions, the \texttt{R} package \texttt{truncnorm} \citep{tnorm2018} must be installed.}
\subsubsection*{design\_analysis( )}
\underline{The function}
\begin{knitrout}\small
\definecolor{shadecolor}{rgb}{0.988, 0.988, 0.988}\color{fgcolor}\begin{kframe}
\begin{verbatim}
> design_analysis( d, n = NULL, power = NULL, sig.level = 0.05,
+                  rangen = c( 2, 1000 ), B = 1e4 , tol=.005 )
\end{verbatim}
\end{kframe}
\end{knitrout}

The function \texttt{design\_analysis( )} runs prospective and retrospective design analysis according to a Cohen's $d$ (\texttt{d}) and a fixed type I error (\texttt{sig.level}). Specifically, if the user specifies:
\begin{itemize}
\item power (\texttt{power}), then it performs \textit{prospective design analysis}
\item sample size per each group (\texttt{n}), then it performs \textit{retrospective design analysis}
\end{itemize}
\textit{Note: It is necessary to provide either \texttt{power} or \texttt{n}.}

\underline{Function arguments}
\begin{itemize}
\item \texttt{d} $=$ Cohen's $d$
\item \texttt{n} $=$ sample size for each group
\item \texttt{power} $=$ power
\item \texttt{sig.level} $=$ type I error. Default is .05
\item \texttt{B} $=$ number of replications for \texttt{d} simulation. Default is 10,000
\item \texttt{rangen} $=$ a vector of two values indicating the range of \texttt{n} within which to search for power.\\
Default is from 2 to 1000.\\
\textit{Note: \texttt{rangen} is used only for prospective design analysis}
\item \texttt{tol} $=$ numerical tolerance used to search power.
Default is .005.\\
\textit{Note: \texttt{tol}  is used only for prospective design analysis}
\end{itemize}

\underline{Function returned values}
\begin{itemize}
\item A list containing main arguments and results
\end{itemize}
\newpage
\underline{Examples}
\begin{singlespace}
\begin{knitrout}\small
\definecolor{shadecolor}{rgb}{0.988, 0.988, 0.988}\color{fgcolor}\begin{kframe}
\begin{verbatim}
> # Prospective design analysis
> design_analysis(d=.50, power=.80, sig.level=.05)
> # $d
> # [1] 0.5
> #
> # $power
> # [1] 0.8
> #
> # $n
> # [1] 64
> #
> # $typeS
> # [1] 0
> #
> # $typeM
> # [1] 1.132944
\end{verbatim}
\end{kframe}
\end{knitrout}
\end{singlespace}
\begin{singlespace}
\begin{knitrout}\small
\definecolor{shadecolor}{rgb}{0.988, 0.988, 0.988}\color{fgcolor}\begin{kframe}
\begin{verbatim}
> # Retrospective design analysis
> design_analysis(d=.50, n=20, sig.level=.05)
> # $d
> # [1] 0.5
> # 
> # $n
> # [1] 20
> # 
> # $power
> # [1] 0.346
> # 
> # $typeS
> # [1] 0.001156069
> # 
> # $typeM
> # [1] 1.739772
\end{verbatim}
\end{kframe}
\end{knitrout}
\end{singlespace}
\newpage
\subsection*{design\_est( )}
\underline{The function}
\begin{knitrout}\small
\definecolor{shadecolor}{rgb}{0.988, 0.988, 0.988}\color{fgcolor}\begin{kframe}
\begin{verbatim}
> design_est( n1 , n2 = n1, target_d = NULL, target_d_limits = NULL, 
+             distribution = c("uniform","normal"), k = 1/6, sig.level = 0.05, 
+             B = 500, B0 = 500, return_data = FALSE )
\end{verbatim}
\end{kframe}
\end{knitrout}
The function \texttt{design\_est( )} performs retrospective design analysis according to a plausible interval  for Cohen's $d$ (see, \texttt{target\_d\_limits}) or to a fixed Cohen's $d$ (see, \texttt{target\_d}) and a fixed type I error (\texttt{sig.level}). Different sample sizes for each sample can be specified.\\ 
\textit{Note: It is necessary to provide either \texttt{target\_d} or \texttt{target\_d\_limits}.}

\underline{Function arguments}
\begin{itemize}
\item \texttt{n1} $=$ sample size of first group
\item \texttt{n2} $=$ sample size of second group. Default is \texttt{n1}
\item \texttt{target\_d} $=$ Cohen's $d$.
\item \texttt{target\_d\_limits} $=$ vector of two values specifing the plausible interval of Cohen's $d$
\item \texttt{distribution} $=$ a character string specifying the probability distribution associated with the plausible interval for Cohen's $d$, must be one of \texttt{"uniform"} or \texttt{"normal"}
\item \texttt{k} $=$ if \texttt{"normal"} is specified as distribution, \texttt{k} is used to define the standard deviation of the doubly truncated normal distribution. Specifically, the standard deviation is calculated as the length of the plausible interval times \texttt{k}. Default is $\frac{1}{6}$
\item \texttt{sig.level} $=$ type I error. Default is .05
\item \texttt{B} $=$ number of replications for Cohen's $d$ simulation. Default is 500
\item \texttt{B0} $=$ number of Cohen's $d$ sampled from the plausible interval. Default is 500
\item \texttt{return\_data} $=$ if \texttt{TRUE} and a plausible interval for Cohen's $d$ is specified, it returns a \texttt{data.frame} of \texttt{B0} rows with power, Type S and Type M errors for each Cohen's $d$ sampled from the plausible interval. Default is \texttt{FALSE}
\end{itemize}

\underline{Function returned values}
Two lists:
\begin{itemize}
\item \texttt{class} containing the main arguments
\item \texttt{results} containing the results
\end{itemize}

\newpage
\underline{Examples}
\begin{singlespace}
\begin{knitrout}\small
\definecolor{shadecolor}{rgb}{0.988, 0.988, 0.988}\color{fgcolor}\begin{kframe}
\begin{verbatim}
> # Retrospective design analysis with plausible interval
> #  and associated doubly truncated normal distribution 
> out <- design_est(n1=50, n2=48, distribution="normal",
+                   target_d_limits = c(.20,.30), return_data=TRUE)
> #
> out
> # $call
> # $call$n1
> # [1] 50
> # 
> # $call$n2
> # [1] 48
> # 
> # $call$target_d_limits
> # [1] 0.2 0.3
> # 
> # $call$distribution
> # [1] "normal"
> # 
> # $call$k
> # [1] 0.1666667
> # 
> # $call$B
> # [1] 500
> # 
> # $call$B0
> # [1] 500
> # 
> # 
> # $results
> # $results$power
> # [1] 0.232588
> # 
> # $results$typeS
> # [1] 0.003919314
> # 
> # $results$typeM
> # [1] 2.090674
> # 
> # $results$data
> #     power       typeS    typeM
> # 1   0.240 0.000000000 2.024220
> # 2   0.162 0.000000000 2.045419
> # 3   0.236 0.000000000 2.104194
> # 4   0.188 0.010638298 2.048976
> # 5   0.208 0.000000000 2.062765
> # ...
> #
> #
> # To draw the distribution of simulated power, Type S and Type M errors:
> hist(out$results$data$power,main="",xlab="power") # power
> hist(out$results$data$typeS,main="",xlab="typeS") # Type S error
> hist(out$results$data$typeM,main="",xlab="typeM") # Type M error
\end{verbatim}
\end{kframe}
\end{knitrout}
\end{singlespace}

\newpage
\subsection*{Examples included in the paper}
\begin{knitrout}\small
\definecolor{shadecolor}{rgb}{0.988, 0.988, 0.988}\color{fgcolor}\begin{kframe}
\begin{verbatim}
> # Clear workspace
> rm(list=ls())
> # Load the functions for design analysis
> source("PRDA.R") 
\end{verbatim}
\end{kframe}
\end{knitrout}

\begin{knitrout}\small
\definecolor{shadecolor}{rgb}{0.988, 0.988, 0.988}\color{fgcolor}\begin{kframe}
\begin{verbatim}
> # Subsection 1.3
> design_analysis(n=33, d=.20) 
\end{verbatim}
\end{kframe}
\end{knitrout}
\begin{knitrout}\small
\definecolor{shadecolor}{rgb}{0.988, 0.988, 0.988}\color{fgcolor}\begin{kframe}
\begin{verbatim}
> # Subsection 2.1
> design_analysis(d=.25, power=.80) 
> design_analysis(d=.25, power=.60) 
\end{verbatim}
\end{kframe}
\end{knitrout}

\begin{knitrout}\small
\definecolor{shadecolor}{rgb}{0.988, 0.988, 0.988}\color{fgcolor}\begin{kframe}
\begin{verbatim}
> # Subsection 2.2
> design_analysis(n=31, d=.25) 
\end{verbatim}
\end{kframe}
\end{knitrout}

\begin{knitrout}\small
\definecolor{shadecolor}{rgb}{0.988, 0.988, 0.988}\color{fgcolor}\begin{kframe}
\begin{verbatim}
> # Section 3
> design_est(n1=31, n2=31, target_d_limits=c(.20,.60), distribution="normal") 
\end{verbatim}
\end{kframe}
\end{knitrout}

\begin{knitrout}\small
\definecolor{shadecolor}{rgb}{0.988, 0.988, 0.988}\color{fgcolor}\begin{kframe}
\begin{verbatim}
> # Section 4: Prospective design analysis
> power1=.80 ; power2=.60
> d1=.20 ; d2=.35 ; d3=.50
> design_analysis(d=d1, power=power1) 
> design_analysis(d=d2, power=power1) 
> design_analysis(d=d3, power=power1) 
> design_analysis(d=d1, power=power2) 
> design_analysis(d=d2, power=power2) 
> design_analysis(d=d3, power=power2) 
> # Section 4: Retrospective design analysis
> design_est(n1=34,n2=33,target_d=d1,B=10000) 
> design_est(n1=34,n2=33,target_d=d2,B=10000) 
> design_est(n1=34,n2=33,target_d=d3,B=10000) 
> # Section 4: Retrospective design analysis with plausible interval
> design_est(n1=34, n2=33, distribution="normal",target_d_limits = c(.25,.45))
\end{verbatim}
\end{kframe}
\end{knitrout}
\vskip.5cm
For any further information please contact:\\
 \url{massimiliano.pastore@ unipd}, \url{gianmarco.altoe@unipd.it}

\end{document}